\documentclass[%
 reprint,
 amsmath,amssymb,
 aps,
]{revtex4-2}

\usepackage{xcolor}
\usepackage{graphicx}% Include figure files
\usepackage{dcolumn}% Align table columns on decimal point
\usepackage{bm}% bold math
\begin{document}

\preprint{APS/123-QED}

\title{Primordial black holes with anisotropic hair}%

\author{Chong-Bin Chen$^{1,2}$}
\author{Bai-Xin An$^{1,2}$}
\author{Fu-Wen Shu$^{1,2}$}%
 \email{shufuwen@ncu.edu.cn}
\affiliation{%
$^{1}$ Department of Physics, Nanchang University, Nanchang, 330031, China\\
$^{2}$ Center for Relativistic Astrophysics and High Energy Physics,
Nanchang University, Nanchang 330031, China
}%


\begin{abstract}
A mechanism for generating anisotropic enhancements of the curvature perturbation through a vector field is proposed. We find that when the mixing between the inflaton perturbation and the vector-field perturbation is sufficiently strong in the anisotropic inflation, the power spectrum becomes dominated by anisotropic constant modes.  This suggests that statistical anisotropy in primordial black hole (PBH) formation may be inevitable if inflation undergoes an anisotropic inflationary phase. Our findings offer a novel approach to probe vector fields during inflation and to test the cosmic no-hair conjecture. 
\end{abstract}


\maketitle

%\tableofcontents

\section{Introduction}

The small-scale physics of inflation remains poorly understood. The cosmic microwave background(CMB) observation only constrains the first few $e$-folds of inflation---yielding an almost scale-invariant, Gaussian and adiabatic curvature perturbation on large scales($k\lesssim1\ \text{Mpc}^{-1}$) \cite{Lyth:1996im,Planck:2018jri}. On small scales, however, the curvature perturbation may be responsible for the creation of PBH with an abundance compatible with dark
matter \cite{Hawking:1971ei,Carr:1974nx,Chapline:1975ojl,Carr:2016drx,Inomata:2017okj}. Such PBH production requires the power spectrum to exhibit a peak(which strongly depends on $k$) on small scales with an amplitude at least $10^7$ times larger than that of the CMB. The enhancement of the amplitude of the curvature perturbation can be generated in several mechanisms \cite{Ozsoy:2023ryl,Jedamzik:1999am,Bugaev:2013fya,Linde:2012bt,Kawasaki:2019hvt,Cai:2018tuh,Green:2000he}.

In this paper, we explore the possibility of large anisotropic curvature perturbation sourced by a vector field and examine their imprint on PBH production when the vector field is amplified through the kinetic coupling $f^2(\phi)F_{\mu\nu}F^{\mu\nu}$, which is originally motivated by supergravity  \cite{Ratra:1991bn}. A broad class of coupling functions permits the so-called anisotropic inflation,  characterized by a constant electric field during inflation \cite{Watanabe:2009ct,Maleknejad:2012fw,Kanno:2010nr,Soda:2012zm,Watanabe:2010bu,Chen:2022ccf}. The key parameter of the phenomenology in anisotropic inflation is $\mathcal{I}$:
\begin{align}\label{I}
    \mathcal{I}\equiv \sqrt{\frac{3\epsilon_A}{2\epsilon_{\phi}}},
\end{align}
which characterizes the mixing between the inflaton perturbation and the vector-field perturbation. Here
$\epsilon_{\phi}$ and $\epsilon_A$ are the velocity of the inflaton and a vector field respectively. The mixing term induces anisotropy in the power spectrum $\mathcal{P}_{\mathcal{R}}$ of the curvature perturbation $\mathcal{R}$. Previous work derived this anisotropic contribution as $\delta\mathcal{P}_{\mathcal{R}}/\mathcal{P}_{\mathcal{R},0}=16\mathcal{I}^2N_k^2\sin^2\theta$ \cite{Watanabe:2010fh,Emami:2013bk,Bartolo:2012sd,Gumrukcuoglu:2007bx,Himmetoglu:2009mk,Gumrukcuoglu:2010yc,Durrer:2024ibi}, where $\mathcal{P}_{\mathcal{R},0}\equiv\mathcal{P}_{\mathcal{R}}(\mathcal{I}=0)$, $N_k\sim60$ represents the $e$-folding number of the mode elapsed since the
horizon-crossing, and $\theta$ is the angle between the vector field and the wave number. The factor $N_k^2$ arises from the cumulative effect of vector-field perturbations acting as a source on superhorizon scales, driving the evolution of $\mathcal{R}$. It seems that we can take $\mathcal{I}\gg1$ of this result to obtain a large enhancement of the power spectrum. However, it has been shown that the in-in perturbation theory breaks down for $\mathcal{I}\gtrsim\mathcal{O}(0.1)$  \footnote{Although \cite{Funakoshi:2012ym,Chen:2023bcz} studied the isotropic configuration of a triad of vector fields, the conclusion should also hold for the case of a single vector field}. The strong mixing of perturbations has been discussed recently in the context of multifield inflation \cite{Garcia-Saenz:2018ifx,Fumagalli:2019noh,Bjorkmo:2019qno,Garcia-Saenz:2019njm,Garcia-Saenz:2018vqf,Garcia-Saenz:2025jis} and is used as a new mechanics of PBH formation \cite{Palma:2020ejf,Fumagalli:2020adf,Braglia:2020eai}. One may concern about applying similar treatments to vector fields.

The purpose of this paper is to investigate the regime $\mathcal{I}\gg1$ of anisotropic inflation \footnote{The isotropic case of a triad of vectors has been studied in \cite{Yamamoto:2012sq,Chen:2022ccf,Chen:2023bcz}. Here we focus on the anisotropic one}. We find that in this regime, the dominant contribution to $\mathcal{R}$ arises not from cumulative-effect modes but from constant modes when $\mathcal{I}\sin\theta\gg1$. To elucidate this mechanism, we will be in the perspective of the effective field theory(EFT) by integrating out heavy modes. Our analysis reveals that the power spectrum exhibits exponential enhancement:
\begin{align}
    \Delta_{\mathcal{R}}\equiv\frac{\mathcal{P}_{\mathcal{R}}}{\mathcal{P}_{\mathcal{R},0}}\sim  e^{4\pi\left|\sin{\theta}-\frac{\sqrt{2}}{2}\right|\mathcal{I}}
\end{align}
for $\sin^2\theta>1/2$. Only modes with $\theta\sim\pi/2$ are sufficiently enhanced, leading to a statistically anisotropic power spectrum. If inflation experiences a transient anisotropic phase, such anisotropic peaks will inevitably imprint on the power spectrum, with potential cosmological implications.

\section{\label{sec:level1}Anisotropic inflation}

We begin by reviewing anisotropic inflation. Consider a model coupling an inflaton field to a vector field through the action
\begin{align}
    S=\int d^4x\sqrt{-g}\bigg[&\frac{M_{\text{pl}}^2}{2}R-\frac{1}{2}\partial^{\mu}\phi\partial_{\mu}\phi-V(\phi)\nonumber\\
    &-\frac{f^2(\phi)}{4}F_{\mu\nu}F^{\mu\nu}\bigg],
\end{align}
where $F_{\mu\nu}\equiv\partial_{\mu}A_{\nu}-\partial_{\nu}A_{\mu}$, and the kinetic function $f(\phi)$ governs the coupling between the inflaton and the vector field. After gauge fixing $A_0=0$, the equation of motion of $A_i(t)$ can be solved as $\dot{A}_i(t)\propto f^{-2}a^{-1}$, where the overdot denoting the derivative with respect to the cosmological time $t$. The energy densities of the electric and magnetic fields are given by \cite{Subramanian:2009fu}
\begin{align}
    \rho_{E}\sim \frac{f^2}{a^2}\dot{A}_i^2,\ \ \ \ \ \ \ \rho_{B}\sim \frac{f^2}{2a^4}\left(\partial_i A_j-\partial_{j}A_{i}\right)^2.
\end{align}
We are interested in a constant electric field that requires $f\sim a^{-2}$. For a homogeneous vector field($\rho_B=0$), the Universe remains spatially homogeneous. However, nonzero $\rho_E$ maintains an anisotropic spacetime during the evolution of the universe. Without loss of generality, adopting the ansatz $A_{\mu}=(0,A(t),0,0)$, the metric of spacetime can be represented as a Bianchi type I form $ds^2=-a^2(\eta)\left(-d\eta^2+dx^2\right)+b^2(\eta)\left(dy^2+dz^2\right)
$, where $\eta$ is the conformal time $d\eta=dt/a$. If the energy density of the vector field remains subdominated compared with inflaton $\rho_A\ll V$, the anisotropy of spacetime is always of the order of slow-roll parameters \cite{Watanabe:2010fh}, i.e., $aH\simeq(-\eta)^{-1-\epsilon_H}$ and $bH\simeq(-\eta)^{-1-\epsilon_H-\epsilon_A}$ so that $b'/b-a'/a\simeq aH\epsilon_A$, where $H\equiv\dot{a}/a$ and the prime denotes the derivative with respect to conformal time $\eta$. The slow-roll parameters are defined by $\epsilon_{\phi}\equiv\dot{\phi}^2/2M_{\text{pl}}^2H^2, \epsilon_A\equiv f^2\dot{A}^2/3a^2M_{\text{pl}}^2H^2$ in our ansatz.

During the anisotropic phase, the potential of inflaton still dominates the inflation, while a portion of its energy transforms to the vector field. In this scenario, both the rolling of the inflaton and the vector field are made responsible for the deviation of the de Sitter space, i.e., $\epsilon_H=\epsilon_{\phi}+\epsilon_{A}$.

A concrete realization of ansiotropic inflation can be achieved by choosing the kinetic coupling as
$f(\phi)=\exp\big(2c\int d\phi\ V/V_{\phi}M_{\text{pl}}^2\big)$ with $c>1$, where $V_{\phi}\equiv dV/d\phi$ \cite{Watanabe:2009ct,Maleknejad:2012fw}. 
%In this configuration, $f=a^{-2c}$ and the energy of the electric field grows as $\rho_E\propto a^{4(c-1)}$. Finally, it reacts back to the dynamics of the inflaton, and 
Inflation can transition into an anisotropic attractor with a constant background electric field \cite{Watanabe:2009ct,Maleknejad:2012fw,Kanno:2010nr,Chen:2022ccf}. The inflaton rolling velocity is then reduced to $\epsilon_{\phi}\simeq \epsilon_{\phi}/c$, and the energy density of the vector field is given by $\rho_E/V\equiv f^2\dot{A}^2/2a^2\simeq \epsilon_H(c-1)/2c$. The parameter $\mathcal{I}$ is fundamentally unconstrained in this background dynamics, with its range $0 < \mathcal{I} < \infty$ determined by $\mathcal{I} \simeq \sqrt{3(c-1)/2}$ in the anisotropic phase. Our analysis will focus particularly on the regime where $\mathcal{I} \gg 1$.

\section{Strong mixing of perturbations}
The vector field breaks the 3D rotational symmetry of the Universe to a 2D rotational one in the $yz$ plane. This symmetry breaking complicates the metric perturbations, with the complete quadratic action having been thoroughly analyzed in previous works \cite{Watanabe:2010fh,Emami:2013bk,Bartolo:2012sd,Gumrukcuoglu:2007bx,Himmetoglu:2009mk,Gumrukcuoglu:2010yc,Durrer:2024ibi}.

However, it is known that the dominated contributions to the curvature perturbation arise only from matter perturbations in the spatially flat gauge, while the interactions between matter and metric are slow-roll suppressed. Consequently, we may safely neglect gravitational back-reaction in the following. The matter perturbations are given by

\begin{align}
    \phi=\phi_0+\varphi, \ \ \ \ \ A_{\mu}^{(S)}=(\delta A_0, A+\delta A_1, 0, 0).
\end{align}
Here, we consider only the scalar parts of the perturbations of $A_\mu$ and have removed d.o.f. by fixing the U(1) symmetry. Without loss of generality, we align the wave number as $\boldsymbol{k}=(k_1,k_2,0)$, where $k_1=k\cos\theta$ and $k_2=(b/a)k\sin\theta$, and $\theta$ being the angle between the vector field and $k$. After eliminating the nondynamical d.o.f. $\delta A_0$ and introducing a new variable $\mathcal{A}\equiv (f/a)\sin\theta\delta A_1$, the dominant quadratic action is given by
\begin{align}\label{action}
    \mathcal{L}_2\simeq \frac{a^3}{2}\Bigg[&|\dot{\varphi}|^2-\left(\frac{k^2}{a^2}+8\mathcal{I}^2H^2\cos{2\theta}\right)|\varphi|^2+|\dot{\mathcal{A}}|^2\nonumber\\
    &-\frac{k^2}{a^2}|\mathcal{A}|^2-4H\mathcal{I}\sin\theta\Big(\varphi^*\dot{\mathcal{A}}+c.c\Big)\nonumber\\
    &-12H^2\mathcal{I}\sin\theta\Big(\varphi^*\mathcal{A}+c.c.\Big) \Bigg].
\end{align}
Here, we have assumed $\dot{\phi}<0$, $\dot{A}>0$, and treat $\mathcal{I}$ as a constant to leading order of the slow-roll approximation. Spacetime anisotropy, being slow-roll suppressed, is also neglected so that $a=b$. 

The last two terms in (\ref{action}) represent the mixing between $\varphi$ and $\mathcal{A}$, which is controlled by $\mathcal{I}\sin\theta$. 
%If $\mathcal{I}\ll 1$ these two perturbations are weakly mixing and we can use in-in formalism to compute the power spectrum of $\varphi$ sourced by the vector perturbation $\mathcal{A}$. This has been well studied in previous work \cite{} and the result depends on the wave number $k$ and angle $\theta$. On the other hand, in the $\mathcal{I}\gtrsim 1$ regime, the mixing is strong for some $\theta$ so that the perturbative method fails. 
We are interested in the strong mixing $\mathcal{I}\sin\theta\gg 1$, which is the
subject of the rest of this paper. 
One can immediately note that in this regime, the mass $m^2_{\varphi}\equiv 8\mathcal{I}^2H^2\cos{2\theta}$ can become heavy for some angle values $\theta$. If this is the case, there is a hierarchy between two frequencies $\omega_{\pm}^2=p^2+4\mathcal{I}^2H^2\pm 4\mathcal{I}H(\mathcal{I}^2H^2+p^2\sin^2\theta)^{1/2}$ of modes on subhorizon scales \cite{Achucarro:2012yr}. Here, $p\equiv k/a$ is the physical momentum. When the momentum of the modes redshifts to $p\ll|m_{\varphi}|$, the high-frequency mode $\omega_+$ can be integrated out, yielding a low-energy effective theory for the light modes \cite{Cheung:2007st,Baumann:2011su,Gwyn:2012mw}.

Specifically, to obtain the leading-order solution for $\varphi$, we denote operator $\square\equiv-\partial^2/\partial t^2-3H\partial/\partial t-k^2/a^2$ and expand the propagator $1/(m^2_{\varphi}-\square)=1/m^2_{\varphi}+\square/m^4_{\varphi}+\cdots$ in the equation of motion. Then we can solve this leading solution as
\begin{align}\label{vaphi_sol}
    \varphi_{\text{LO}}=-\frac{\sin\theta\big(\dot{\mathcal{A}}+3H\mathcal{A}\big)}{2\mathcal{I}H\cos{2\theta}}
\end{align}
for $|m_{\varphi}^2|\gg H^2$. The heavy d.o.f can be integrated out by substituting this solution into the quadratic action. Then the leading-order action of the effective theory of $\mathcal{A}$ is given by
\begin{align}
    \mathcal{L}_{2, \text{LO}}\simeq \frac{a^3}{2c_s^2(\theta)}\left[|\dot{\mathcal{A}}|^2-\frac{c_s^2(\theta)k^2}{a^2}|\mathcal{A}|^2\right],
\end{align}
where the sound speed is given by
\begin{align}
    c_s^2(\theta)=\cos{2\theta}.
\end{align}
This is an effective action of a massless perturbation with sound speed $c_s$. Remarkably, $c_s$ becomes imaginary for $\theta\in(\pi/4,3\pi/4)\cup(5\pi/4,7\pi/4)$, leading to exponentially growing solutions before the horizon crossing \footnote{Recently, the multi-field inflation with strongly non-geodesic attractors in the field space has been extensively explored \cite{Brown:2017osf,Bjorkmo:2019fls}. The effective theory of inflationary dynamics is a massless mode with imaginary sound speed \cite{Garcia-Saenz:2018ifx,Fumagalli:2019noh,Bjorkmo:2019qno,Garcia-Saenz:2019njm,Garcia-Saenz:2018vqf,Garcia-Saenz:2025jis}. It can also provide the mechanism for generating PBH in the multi-filed scenario \cite{Palma:2020ejf,Fumagalli:2020adf}}. Consequently, the power spectrum of $\mathcal{A}$ is scale invariant(of leading order) and enhanced on large scales. This enhancement is anisotropic due to the angle-dependent sound speed $c_s(\theta)$. On the other hand, if $\sin\theta\simeq0$, $\varphi$ and $\mathcal{A}$ are decoupled and $\varphi$ is also heavy with a real mass and decay away on the superhorizon for any value of large $\mathcal{I}$.

The enhanced solution for $\mathcal{A}$ on the superhorizon can be found by solving the coupled system inside the horizon \cite{Bjorkmo:2019qno,Christodoulidis:2023eiw,Chen:2023bcz}. Here, we are only interested in the enhanced solution, where $\sin^2{\theta}>1/2$. The conjugate momentum of the perturbations is given by
\begin{align}
  \pi_{\varphi}=\frac{a^3}{2}\dot{\varphi}, \ \ \ \ \ \pi_{\mathcal{A}}=\frac{a^3}{2}\left(\dot{\mathcal{A}}-4\mathcal{I}\sin{\theta}H\varphi\right).
\end{align}
After defining $p_{\varphi}\equiv2\pi_{\varphi}/a^3H$, $p_{\mathcal{A}}\equiv2\pi_{\mathcal{A}}/a^3H$, $x\equiv-k\eta$\, and using the $e$-folding number as the time variable, the coupled system (\ref{action}) gives $X_{,N}=J_{0}X+J_{1}X$, where $X=(\mathcal{A},p_{\mathcal{A}},\varphi,p_{\varphi})^{T}$, 
\begin{align}
  J_{0}=&\left(\
  \begin{matrix}
  0 & 1 & 4\mathcal{I}\sin{\theta} & 0 &\\
  -x^2 & 0 & 0 & 0 &\\
  0 & 0 & 0 & 0 &\\
  0 & -4\mathcal{I}\sin{\theta} & -(x^2+8\mathcal{I}^2) & 0&
\end{matrix}
\right), \nonumber\\
  J_{1}=&\left(\
  \begin{matrix}
  0 & 0 & 0 & 0 &\\
  0 & -3 & -12\mathcal{I}\sin{\theta} & 0 &\\
  0 & 0 & 0 & 0 &\\
  -12\mathcal{I}\sin{\theta} & 0 & 0 & -3&
\end{matrix}
\right).
\end{align}
Considering the Hubble term as a small correction to the system on subhorizon, the matrix $J_{0}$ is of $0th$ order while $J_{1}$ is first order. Near the horizon crossing, the exponential growth mode of $\mathcal{A}$ emerges from the imaginary frequency $\omega_{-}$ of the coupled system when $p<|m_{\varphi}|$, which yields the enhanced solution at the limit $-k\eta\to0$ \cite{supp},
\begin{align}\label{Asol}
\mathcal{A}\simeq \left|2-4\sin^2\theta\right|^{1/4}e^{2\pi\left|\sin{\theta}-\frac{\sqrt{2}}{2}\right|\mathcal{I}}.
\end{align}
The enhanced factor is, as expected from the sound speed of the EFT, dependent on the angle $\theta$.

One thing to mention is that even on superhorizon scales, $\varphi$ and $\mathcal{A}$ remain coupled for $\sin\theta\neq0$. This can be realized from solution (\ref{vaphi_sol}) that when $\dot{\mathcal{A}}=0$ on the superhorizon, the two perturbations are still totally relevant, i.e., 
\begin{align}\label{relevant}
    \varphi_{\text{LO}}=-\frac{3\sin\theta}{2\mathcal{I}\cos{2\theta}}\mathcal{A},\ \ \ \ \left(-k\eta\ll 1\right)
\end{align}
We found that in this limit, the perturbation $\mathcal{A}$ is dominates over the perturbation $\varphi$. This persistent coupling implies that if anisotropic inflation with $\mathcal{I}\gg1$ occurs after the time that the CMB modes exit the horizon, these large-scale modes can still be kicked by the vector field.

\section{Seeding PBH from a vector field}
\subsection{Features of the power spectrum}
The PBH can be formed if the curvature perturbation generated on small scales during inflation is enhanced. For an anisotropic inflationary universe, the uniform-energy curvature perturbation we can use in the spatially flat gauge is
\begin{align}
    \zeta\simeq-H\frac{\delta\rho}{\dot{\rho}}.
\end{align}
One can directly compute the perturbation of the total energy density of matter to get the curvature perturbation. However, alternatively, here we compute a variable:
\begin{align}\label{AnisoR}
    \mathcal{R}= H\frac{ik^a}{k^2}V_a
\end{align}
in the spatially flat gauge, where $a=1,2,3$ and the total velocity $V_a\equiv(\rho+P)^{-1}T^t_{\ a}$. This reduces to the standard comoving curvature perturbation $\mathcal{R}_{\text{iso}}=-H(\rho+P)u$ in isotropic backgrounds, where $u$ is the scalar velocity of fluids $v_a=\partial_au+u_a$. One can find that in our configuration of anisotropic inflation, $\mathcal{R}=-\zeta$ holds on superhorizon scales up to slow-roll corrections from spacetime anisotropy \cite{supp}. We therefore use $\mathcal{R}$ to compute the power spectrum, consistent with our treatment of neglecting slow-roll suppressed terms.

The curvature perturbation of the anisotropic inflation is then given by
\begin{align}\label{R}
    \mathcal{R}=\frac{1}{\sqrt{2\epsilon_H}M_{\text{pl}}H}\frac{-\varphi+\mathcal{I}\sin\theta\mathcal{A}}{\sqrt{1+2\mathcal{I}^2/3}}.
\end{align}
In the EFT theory of $\mathcal{A}$ (\ref{relevant}), the dominant contribution to the curvature perturbation is from the perturbation of the vector field $\mathcal{A}$ due to the contributions in (\ref{R}) of two perturbations $\sim \mathcal{I}^2\gg 1$. If we assume that $\epsilon_H$ is a constant during inflation, we have
\begin{align}
    \mathcal{R}\propto \sqrt{\frac{3}{2}}\sin{\theta}\mathcal{A}.
\end{align}
The Fourier modes of $\mathcal{R}$ can be decomposed in terms of two sets of operators since we have two perturbations: $\tilde{\mathcal{R}}(\boldsymbol{k},\eta)=\sum_{\alpha=\pm}\left[\mathcal{R_{\alpha}}(k,\eta)a_{\alpha}(\boldsymbol{k})+\mathcal{R^*_{\alpha}}(k,\eta)a_{\alpha}^{\dagger}(-\boldsymbol{k})\right]$, where $a_{\alpha}^{\dagger}(\boldsymbol{k})$ and $a_{\alpha}(\boldsymbol{k})$ are creation and annihilation operators that satisfy the commutation relation $[a_{\alpha}(\boldsymbol{k}),a_{\beta}^{\dagger}(\boldsymbol{k'})]=(2\pi)^3\delta_{\alpha\beta}\delta^{(3)}(\boldsymbol{k}-\boldsymbol{k'})$. The dimensionless power spectrum of the curvature perturbations is then computed by
\begin{align}
    \mathcal{P}_{\mathcal{R}}(k)=\frac{k^3}{2\pi^2}\sum_{\alpha=\pm}|\mathcal{R}_{\alpha}(k)|^2.
\end{align}

%We can solve the coupled e.o.m of the perturbations to get (\ref{R}). The e.o.m of the mode functions are obtained from the quadratic action
%\begin{align}
%    &\bar{\varphi}_{\alpha}''+\left(k^2-\frac{2-8\mathcal{I}^2(1-2\sin^2\theta)}{\eta^2}\right)\bar{\varphi}_{\alpha}=\frac{4\mathcal{I}\sin\theta}{\eta}\bar{\mathcal{A}}_{\alpha}'-\frac{8\mathcal{I}\sin\theta}{\eta^2}\bar{\mathcal{A}}_{\alpha}\nonumber\\
 %   &\bar{\mathcal{A}}_{\alpha}''+\left(k^2-\frac{2}{\eta^2}\right)\bar{\mathcal{A}}_{\alpha}=-\frac{4\mathcal{I}\sin\theta}{\eta}\bar{\varphi}_{\alpha}'-\frac{4\mathcal{I}\sin\theta}{\eta^2}\bar{\varphi}_{\alpha},
%\end{align}
%where $\bar{\varphi}\equiv a\varphi$ and $\bar{\mathcal{A}}\equiv a\mathcal{A}$ are the canonical variables. 

Here, we consider a transient anisotropic phase, i.e., $\mathcal{I}\gg1$, in a period $\Delta N\gtrsim \log \mathcal{I}$. We assume that $\mathcal{I}$ reaches its maximum value $\mathcal{I}(N_f)=\mathcal{I}_{\text{max}}$ at time $N_f>N_{\text{CMB}}$, to see how a time-dependent $\mathcal{I}(\eta)$ can result in an anisotropic peak of the curvature power spectrum on small scales \footnote{We assume that at time $N\ll N_f$ and $N\gg N_f$, inflation is in the anisotropic phase with $\mathcal{I}\ll 1$ so that the vector field $\mathcal{A}$ is approximately a massless field.}. The momentum corresponding to modes that cross the horizon at time $N_f$ is defined as $k_f\equiv a(N_f)H$. In this case, only modes with wave number $k\sim k_f$ can experience exponential growth when crossing the horizon. The power spectrum of the enhanced curvature perturbation near $k\sim k_f$ can be estimated as
\begin{align}
    \Delta_{\mathcal{R}} \simeq \left(\frac{3}{2}\sin^2{\theta} \right)\mathcal{P}_{\mathcal{A}}(k),
\end{align}
where $\mathcal{P}_{\mathcal{A}}=\langle\mathcal{A}\mathcal{A}\rangle$. As is known from the solution of $\mathcal{A}$ (\ref{Asol}), the power spectrum depends exponentially on $\sin\theta$.  The maximum enhancement is at $\theta=\pi/2$. Even a slight deviation $\Delta\theta\sim\mathcal{I}^{-1}$ from $\pi/2$ leads to a drastic suppression of the peak of the power spectrum. 

On the other hand, we define the mass-crossing time $k/a(\tilde{N})=|m_{\varphi}(\tilde{N})|$ for any $k$ mode because most of the growth occurs at the early time \cite{Bjorkmo:2019qno,Christodoulidis:2023eiw,Fumagalli:2020adf}. We denote the location of the peak of the power spectrum as $k_p$. $k_p$ is then determined by $\tilde{N}\sim N_f$; i.e., the mode that crosses the mass-horizon at time $N_f$,
\begin{align}
    k_p(\theta)\sim \frac{|m_{\varphi}(\theta)|}{H}\cdot k_f,
\end{align}
which means that the location of the peaks is also anisotropic.

Modes with $k\gg k_f$ experience nothing because the vector field has been turned off well before horizon crossing. On the other hand, large-scale modes $k\gg k_f$ are still almost scale invariant but are kicked by the vector, which results in a shift deformation. The amplitudes of the shift obviously depend on the angle $\theta$, and the large-scale power spectrum is significantly anisotropy. This is inconsistent with the CMB observation. However, this work presents one minimal mechanism for generating anisotropic peaks of the power spectrum at small scales. The large-scale power spectrum may originate from alternative mechanisms, such as the curvaton scenario, depending upon the specific model framework. We provide a toy example in the following.

\subsection{A toy example}
We consider a model containing two scalar fields and a vector field $\mathcal{L}=\mathcal{L}_{\phi\text{-}A}+\mathcal{L}_{\psi}$, where
\begin{align}
    \mathcal{L}_{\phi\text{-}A}&=\sqrt{-g}\bigg[-\frac{1}{2}\partial^{\mu}\phi\partial_{\mu}\phi-V(\phi)-\frac{f^2(\phi)}{4}F_{\mu\nu}F^{\mu\nu}\bigg],\nonumber\\
    \mathcal{L}_{\psi}&=\sqrt{-g}\bigg[-\frac{1}{2}\partial^{\mu}\psi\partial_{\mu}\psi-U(\psi)\bigg].
\end{align}
$\mathcal{L}_{\psi}$ is only coupled to $\mathcal{L}_{\phi\text{-}A}$ through gravitational interaction. Hence, the power spectrum of the curvature perturbation of this system is given by
\begin{align}\label{toyP}
    \mathcal{P}_{\mathcal{R}}(k)&\simeq\mathcal{P}_{\psi}(k)+\mathcal{P}_{\phi\text{-}A}(k)\nonumber\\
    &\propto\frac{r^2|\delta\psi|^2+|-\varphi+\mathcal{I}\sin{\theta\mathcal{A}}|^2}{r^2+1+2\mathcal{I}^2/3},
\end{align}
where $r\equiv|\dot{\psi}/\dot{\phi}|$. 

\begin{figure}[th]
\centering
\includegraphics[width=0.4\textwidth]{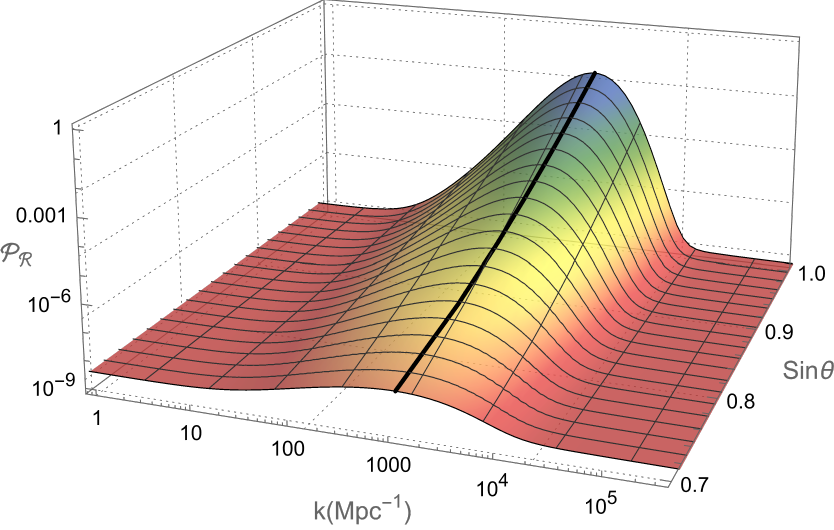}
\caption{\label{fig:PS3D}The power spectrum of $\mathcal{R}$ for our toy example (\ref{toyP}). We use the Gaussian profile $\mathcal{I}(N)=\mathcal{I}_{\text{max}} e^{-(N-N_f)^2/(2\Delta N^2)}$ with $\mathcal{I}_{\text{max}}=5$, $\Delta N=2$, $N_f\simeq N_{\text{CMB}}+5$ and $r_{\text{ini}}=10$. $\mathcal{P}_{\mathcal{A}}=\mathcal{P}_0\left|2-4\sin^2\theta\right|^{1/2}e^{4\pi\left|\sin{\theta}-\frac{\sqrt{2}}{2}\right|\mathcal{I}}\big{|}_{N_v(k)}$, where  $N_v(k)$ is given by $p(N_v)=|m_{\varphi}(N_v)|$. The black curve represent peaks $k_p(\theta)$ of $\mathcal{P}_{\mathcal{R}}$.}
\end{figure}

In the early stage of inflation, the energy density of $A$ is negligible, reducing the system to a two-field system. Assuming slow-roll conditions for both $\psi$ and $\phi$, adiabatic and entropy perturbations remain decoupled at early times \cite{Gordon:2000hv}. Consequently, the curvature perturbation remains conserved on large scales. The ratio $r$ evolves as $r(t)=r_{\text{ini}}[1+2\mathcal{I}(t)^2/3]$ due to the anisotropic phase of the late time. We assume that $r_{\text{ini}}\gg1$. In this regime, $\mathcal{P}_{\mathcal{R}}\simeq\mathcal{P}_{\psi}=(H/\dot{\psi})^2|\delta\psi|^2=2.1\times 10^{-9}$ dominates the power spectrum on large scales($k\lesssim 1\  \text{Mpc}^{-1}$). At time $N_f$ after the time in which the CMB modes exit the horizon, $\mathcal{I}$ becomes significant. Then the power spectrum is dominated by the $\mathcal{P}_{\phi\text{-}A}$ on scale $k_f$. One of the examples of this model is shown in Fig. \ref{fig:PS3D}. We can obviously observe the anisotropic enhancements and the anisotropic peaks(the black curve).

\section{Conclusion}
We have shown that anisotropic inflation allows for an exponential enhancement in the power spectrum for $\mathcal{I}\gg1$. This work presents the first fully analytical treatment of the constant, large $\mathcal{I}$ regime, revealing fundamentally different behavior from previous studies based on a small $\mathcal{I}$. Crucially, we find that superhorizon power spectrum contributions are dominated by enhanced constant modes.

The most interesting feature of the enhanced power spectrum sourced from the vector itself is the statistical anisotropy. Within the EFT framework, the light field $\mathcal{A}$ develops a $\theta$-dependent sound speed $c_s^2=\cos2\theta$ during horizon crossing. Notably, when $c_s^2<0$, the imaginary sound speed triggers exponential growth of the light field. 

This mechanism is not restricted to specific models—it can be triggered whenever the choices of $f(\phi)$ and $V(\phi)$ drive inflation into an anisotropic phase. Therefore, once inflation enters an anisotropic phase at a certain time $N_f$, it produces a corresponding anisotropic peak in the power spectrum at scale $k_p$. However, several aspects require careful consideration. We need to future study their non-Gaussianity(NG) and perturbative control. One may be concerned that the exponential enhancement of perturbations leads to an incredible enhancement on NG. However, it has been shown that the low-energy EFT with imaginary sound speed reduced from strongly nongeodesic attractors of multifield inflation generates only mild NG \cite{Bjorkmo:2019qno,Fumagalli:2019noh,Garcia-Saenz:2019njm,Garcia-Saenz:2018vqf}. Moreover, the perturbative validity is also well controlled in this attractor \cite{Fumagalli:2019noh,Bjorkmo:2019qno,Garcia-Saenz:2025jis}. We believe that this analysis can be used in our case in this paper. Moreover, we only consider $\Delta N\gtrsim \log\mathcal{I}$ in this paper. References \cite{Palma:2020ejf,Fumagalli:2020adf} demonstrate that the curvature perturbation power spectrum exhibits oscillatory features when $\Delta N\lesssim \log\mathcal{I}$. It is interesting to explore this case under our anisotropic model. The anisotropic hair may imprint distinctive features in this power spectrum.

The anisotropic features of the peak can be caught by some future observations, such as PBH production. As we know, the PBH production of a statistically anisotropic peak is still absent. It is necessary to figure out the physical details of this case. In particular, anisotropic hair may be imprinted on the initial spin of PBHs, leaving behind intriguing observational signatures \cite{Harada:2024jxl}. Moreover, enhancements of the curvature power spectrum on small scales are generally accompanied by the production of scalar-induced gravitational waves(SIGWs). This implies that anisotropic features will also manifest as sources in the SIGW spectrum. 
The frequencies of SIGWs is related to $N_f$ as $N_f+\log\left(\mathcal{I}\right)\sim \log\left(f_{\text{peak}}/\text{Hz}\right)+37$ \cite{Fumagalli:2020nvq}, which spans a broad spectrum covered by gravitational-wave experiments, such as PTA and LISA. A model-independent discussion of SIGWs sourced from a statistically anisotropic scalar spectrum has been studied in \cite{Chen:2022qec,Kuang:2023urj}. We provide a specific realization of a statistically anisotropic peak in this paper. Moreover, the scalar modes can be the source of GWs generated during inflation. The GWs are enhanced by the presence of a scalar excited state and overcome the SIGWs \cite{Fumagalli:2021mpc}. It is interesting to explore signatures of this anisotropic hair in the stochastic gravitational wave background(SGWB) and test it in future observations.

\begin{acknowledgments}
We would like to thank Jiro Soda for helpful discussion. This work was supported by the National Natural Science Foundation of China with the Grant No. 12375049, and Key Program of the Natural Science Foundation of Jiangxi Province under Grant No. 20232ACB201008.
\end{acknowledgments}

\nocite{*}

\bibliography{Aniso_PBH_ref}% Produces the bibliography via BibTeX.

\providecommand{\noopsort}[1]{}\providecommand{\singleletter}[1]{#1}%
\begin{thebibliography}{12}%
\makeatletter
\providecommand \@ifxundefined [1]{%
 \@ifx{#1\undefined}
}%
\providecommand \@ifnum [1]{%
 \ifnum #1\expandafter \@firstoftwo
 \else \expandafter \@secondoftwo
 \fi
}%
\providecommand \@ifx [1]{%
 \ifx #1\expandafter \@firstoftwo
 \else \expandafter \@secondoftwo
 \fi
}%
\providecommand \natexlab [1]{#1}%
\providecommand \enquote  [1]{``#1''}%
\providecommand \bibnamefont  [1]{#1}%
\providecommand \bibfnamefont [1]{#1}%
\providecommand \citenamefont [1]{#1}%
\providecommand \href@noop [0]{\@secondoftwo}%
\providecommand \href [0]{\begingroup \@sanitize@url \@href}%
\providecommand \@href[1]{\@@startlink{#1}\@@href}%
\providecommand \@@href[1]{\endgroup#1\@@endlink}%
\providecommand \@sanitize@url [0]{\catcode `\\12\catcode `\$12\catcode
  `\&12\catcode `\#12\catcode `\^12\catcode `\_12\catcode `\%12\relax}%
\providecommand \@@startlink[1]{}%
\providecommand \@@endlink[0]{}%
\providecommand \url  [0]{\begingroup\@sanitize@url \@url }%
\providecommand \@url [1]{\endgroup\@href {#1}{\urlprefix }}%
\providecommand \urlprefix  [0]{URL }%
\providecommand \Eprint [0]{\href }%
\providecommand \doibase [0]{https://doi.org/}%
\providecommand \selectlanguage [0]{\@gobble}%
\providecommand \bibinfo  [0]{\@secondoftwo}%
\providecommand \bibfield  [0]{\@secondoftwo}%
\providecommand \translation [1]{[#1]}%
\providecommand \BibitemOpen [0]{}%
\providecommand \bibitemStop [0]{}%
\providecommand \bibitemNoStop [0]{.\EOS\space}%
\providecommand \EOS [0]{\spacefactor3000\relax}%
\providecommand \BibitemShut  [1]{\csname bibitem#1\endcsname}%
\let\auto@bib@innerbib\@empty
%</preamble>
\bibitem [{\citenamefont {Subramanian}(2010)}]{Subramanian:2009fu}%
  \BibitemOpen
  \bibfield  {author} {\bibinfo {author} {\bibfnamefont {K.}~\bibnamefont
  {Subramanian}},\ }\bibfield  {title} {\bibinfo {title} {{Magnetic fields in
  the early universe}},\ }\href {https://doi.org/10.1002/asna.200911312}
  {\bibfield  {journal} {\bibinfo  {journal} {Astron. Nachr.}\ }\textbf
  {\bibinfo {volume} {331}},\ \bibinfo {pages} {110} (\bibinfo {year}
  {2010})},\ \Eprint {https://arxiv.org/abs/0911.4771} {arXiv:0911.4771
  [astro-ph.CO]} \BibitemShut {NoStop}%
\bibitem [{\citenamefont {Watanabe}\ \emph {et~al.}(2010)\citenamefont
  {Watanabe}, \citenamefont {Kanno},\ and\ \citenamefont
  {Soda}}]{Watanabe:2010fh}%
  \BibitemOpen
  \bibfield  {author} {\bibinfo {author} {\bibfnamefont {M.-a.}\ \bibnamefont
  {Watanabe}}, \bibinfo {author} {\bibfnamefont {S.}~\bibnamefont {Kanno}},\
  and\ \bibinfo {author} {\bibfnamefont {J.}~\bibnamefont {Soda}},\ }\bibfield
  {title} {\bibinfo {title} {{The Nature of Primordial Fluctuations from
  Anisotropic Inflation}},\ }\href {https://doi.org/10.1143/PTP.123.1041}
  {\bibfield  {journal} {\bibinfo  {journal} {Prog. Theor. Phys.}\ }\textbf
  {\bibinfo {volume} {123}},\ \bibinfo {pages} {1041} (\bibinfo {year}
  {2010})},\ \Eprint {https://arxiv.org/abs/1003.0056} {arXiv:1003.0056
  [astro-ph.CO]} \BibitemShut {NoStop}%
\bibitem [{\citenamefont {Watanabe}\ \emph {et~al.}(2009)\citenamefont
  {Watanabe}, \citenamefont {Kanno},\ and\ \citenamefont
  {Soda}}]{Watanabe:2009ct}%
  \BibitemOpen
  \bibfield  {author} {\bibinfo {author} {\bibfnamefont {M.-a.}\ \bibnamefont
  {Watanabe}}, \bibinfo {author} {\bibfnamefont {S.}~\bibnamefont {Kanno}},\
  and\ \bibinfo {author} {\bibfnamefont {J.}~\bibnamefont {Soda}},\ }\bibfield
  {title} {\bibinfo {title} {{Inflationary Universe with Anisotropic Hair}},\
  }\href {https://doi.org/10.1103/PhysRevLett.102.191302} {\bibfield  {journal}
  {\bibinfo  {journal} {Phys. Rev. Lett.}\ }\textbf {\bibinfo {volume} {102}},\
  \bibinfo {pages} {191302} (\bibinfo {year} {2009})},\ \Eprint
  {https://arxiv.org/abs/0902.2833} {arXiv:0902.2833 [hep-th]} \BibitemShut
  {NoStop}%
\bibitem [{\citenamefont {Maleknejad}\ \emph {et~al.}(2013)\citenamefont
  {Maleknejad}, \citenamefont {Sheikh-Jabbari},\ and\ \citenamefont
  {Soda}}]{Maleknejad:2012fw}%
  \BibitemOpen
  \bibfield  {author} {\bibinfo {author} {\bibfnamefont {A.}~\bibnamefont
  {Maleknejad}}, \bibinfo {author} {\bibfnamefont {M.~M.}\ \bibnamefont
  {Sheikh-Jabbari}},\ and\ \bibinfo {author} {\bibfnamefont {J.}~\bibnamefont
  {Soda}},\ }\bibfield  {title} {\bibinfo {title} {{Gauge Fields and
  Inflation}},\ }\href {https://doi.org/10.1016/j.physrep.2013.03.003}
  {\bibfield  {journal} {\bibinfo  {journal} {Phys. Rept.}\ }\textbf {\bibinfo
  {volume} {528}},\ \bibinfo {pages} {161} (\bibinfo {year} {2013})},\ \Eprint
  {https://arxiv.org/abs/1212.2921} {arXiv:1212.2921 [hep-th]} \BibitemShut
  {NoStop}%
\bibitem [{\citenamefont {Emami}\ and\ \citenamefont
  {Firouzjahi}(2013)}]{Emami:2013bk}%
  \BibitemOpen
  \bibfield  {author} {\bibinfo {author} {\bibfnamefont {R.}~\bibnamefont
  {Emami}}\ and\ \bibinfo {author} {\bibfnamefont {H.}~\bibnamefont
  {Firouzjahi}},\ }\bibfield  {title} {\bibinfo {title} {{Curvature
  Perturbations in Anisotropic Inflation with Symmetry Breaking}},\ }\href
  {https://doi.org/10.1088/1475-7516/2013/10/041} {\bibfield  {journal}
  {\bibinfo  {journal} {JCAP}\ }\textbf {\bibinfo {volume} {10}},\ \bibinfo
  {pages} {041}},\ \Eprint {https://arxiv.org/abs/1301.1219} {arXiv:1301.1219
  [hep-th]} \BibitemShut {NoStop}%
\bibitem [{\citenamefont {Bartolo}\ \emph {et~al.}(2013)\citenamefont
  {Bartolo}, \citenamefont {Matarrese}, \citenamefont {Peloso},\ and\
  \citenamefont {Ricciardone}}]{Bartolo:2012sd}%
  \BibitemOpen
  \bibfield  {author} {\bibinfo {author} {\bibfnamefont {N.}~\bibnamefont
  {Bartolo}}, \bibinfo {author} {\bibfnamefont {S.}~\bibnamefont {Matarrese}},
  \bibinfo {author} {\bibfnamefont {M.}~\bibnamefont {Peloso}},\ and\ \bibinfo
  {author} {\bibfnamefont {A.}~\bibnamefont {Ricciardone}},\ }\bibfield
  {title} {\bibinfo {title} {{Anisotropic power spectrum and bispectrum in the
  $f(\phi)F^2$ mechanism}},\ }\href
  {https://doi.org/10.1103/PhysRevD.87.023504} {\bibfield  {journal} {\bibinfo
  {journal} {Phys. Rev. D}\ }\textbf {\bibinfo {volume} {87}},\ \bibinfo
  {pages} {023504} (\bibinfo {year} {2013})},\ \Eprint
  {https://arxiv.org/abs/1210.3257} {arXiv:1210.3257 [astro-ph.CO]}
  \BibitemShut {NoStop}%
\bibitem [{Note1()}]{Note1}%
  \BibitemOpen
  \bibinfo {note} {We assume that at time $N\ll N_f$ and $N\gg N_f$, inflation
  is in the anisotropic phase with $\protect \mathcal {I}\ll 1$ so that the
  gauge field $\protect \mathcal {A}$ is approximately a massless
  field.}\BibitemShut {Stop}%
\bibitem [{\citenamefont {Bjorkmo}\ \emph {et~al.}(2019)\citenamefont
  {Bjorkmo}, \citenamefont {Ferreira},\ and\ \citenamefont
  {Marsh}}]{Bjorkmo:2019qno}%
  \BibitemOpen
  \bibfield  {author} {\bibinfo {author} {\bibfnamefont {T.}~\bibnamefont
  {Bjorkmo}}, \bibinfo {author} {\bibfnamefont {R.~Z.}\ \bibnamefont
  {Ferreira}},\ and\ \bibinfo {author} {\bibfnamefont {M.~C.~D.}\ \bibnamefont
  {Marsh}},\ }\bibfield  {title} {\bibinfo {title} {{Mild Non-Gaussianities
  under Perturbative Control from Rapid-Turn Inflation Models}},\ }\href
  {https://doi.org/10.1088/1475-7516/2019/12/036} {\bibfield  {journal}
  {\bibinfo  {journal} {JCAP}\ }\textbf {\bibinfo {volume} {12}},\ \bibinfo
  {pages} {036}},\ \Eprint {https://arxiv.org/abs/1908.11316} {arXiv:1908.11316
  [hep-th]} \BibitemShut {NoStop}%
\bibitem [{\citenamefont {Christodoulidis}\ and\ \citenamefont
  {Gong}(2024)}]{Christodoulidis:2023eiw}%
  \BibitemOpen
  \bibfield  {author} {\bibinfo {author} {\bibfnamefont {P.}~\bibnamefont
  {Christodoulidis}}\ and\ \bibinfo {author} {\bibfnamefont {J.-O.}\
  \bibnamefont {Gong}},\ }\bibfield  {title} {\bibinfo {title} {{Enhanced power
  spectra from multi-field inflation}},\ }\href
  {https://doi.org/10.1088/1475-7516/2024/08/062} {\bibfield  {journal}
  {\bibinfo  {journal} {JCAP}\ }\textbf {\bibinfo {volume} {08}},\ \bibinfo
  {pages} {062}},\ \Eprint {https://arxiv.org/abs/2311.04090} {arXiv:2311.04090
  [hep-th]} \BibitemShut {NoStop}%
\bibitem [{\citenamefont {Fumagalli}\ \emph {et~al.}(2023)\citenamefont
  {Fumagalli}, \citenamefont {Renaux-Petel}, \citenamefont {Ronayne},\ and\
  \citenamefont {Witkowski}}]{Fumagalli:2020adf}%
  \BibitemOpen
  \bibfield  {author} {\bibinfo {author} {\bibfnamefont {J.}~\bibnamefont
  {Fumagalli}}, \bibinfo {author} {\bibfnamefont {S.}~\bibnamefont
  {Renaux-Petel}}, \bibinfo {author} {\bibfnamefont {J.~W.}\ \bibnamefont
  {Ronayne}},\ and\ \bibinfo {author} {\bibfnamefont {L.~T.}\ \bibnamefont
  {Witkowski}},\ }\bibfield  {title} {\bibinfo {title} {{Turning in the
  landscape: A new mechanism for generating primordial black holes}},\ }\href
  {https://doi.org/10.1016/j.physletb.2023.137921} {\bibfield  {journal}
  {\bibinfo  {journal} {Phys. Lett. B}\ }\textbf {\bibinfo {volume} {841}},\
  \bibinfo {pages} {137921} (\bibinfo {year} {2023})},\ \Eprint
  {https://arxiv.org/abs/2004.08369} {arXiv:2004.08369 [hep-th]} \BibitemShut
  {NoStop}%
\bibitem [{\citenamefont {Gordon}\ \emph {et~al.}(2000)\citenamefont {Gordon},
  \citenamefont {Wands}, \citenamefont {Bassett},\ and\ \citenamefont
  {Maartens}}]{Gordon:2000hv}%
  \BibitemOpen
  \bibfield  {author} {\bibinfo {author} {\bibfnamefont {C.}~\bibnamefont
  {Gordon}}, \bibinfo {author} {\bibfnamefont {D.}~\bibnamefont {Wands}},
  \bibinfo {author} {\bibfnamefont {B.~A.}\ \bibnamefont {Bassett}},\ and\
  \bibinfo {author} {\bibfnamefont {R.}~\bibnamefont {Maartens}},\ }\bibfield
  {title} {\bibinfo {title} {{Adiabatic and entropy perturbations from
  inflation}},\ }\href {https://doi.org/10.1103/PhysRevD.63.023506} {\bibfield
  {journal} {\bibinfo  {journal} {Phys. Rev. D}\ }\textbf {\bibinfo {volume}
  {63}},\ \bibinfo {pages} {023506} (\bibinfo {year} {2000})},\ \Eprint
  {https://arxiv.org/abs/astro-ph/0009131} {arXiv:astro-ph/0009131}
  \BibitemShut {NoStop}%
\bibitem [{\citenamefont {Noh}\ and\ \citenamefont {Hwang}(1995)}]{Noh:1987vk}%
  \BibitemOpen
  \bibfield  {author} {\bibinfo {author} {\bibfnamefont {H.}~\bibnamefont
  {Noh}}\ and\ \bibinfo {author} {\bibfnamefont {J.~C.}\ \bibnamefont
  {Hwang}},\ }\bibfield  {title} {\bibinfo {title} {{Perturbations of an
  anisotropic space-time: Formulation}},\ }\href
  {https://doi.org/10.1103/PhysRevD.52.1970} {\bibfield  {journal} {\bibinfo
  {journal} {Phys. Rev. D}\ }\textbf {\bibinfo {volume} {52}},\ \bibinfo
  {pages} {1970} (\bibinfo {year} {1995})}\BibitemShut {NoStop}%
\end{thebibliography}%

\end{document}

\title{Supplemental Material}

\maketitle

\section{Enhanced modes with constant $\mathcal{I}$}\label{app:modesolution}
One can solve the coupled system in the subhorizon $k/aH\gg 1$ to find an enhanced solution by considering the Hubble friction as a first-order correction. This was recently studied in the rapid-turn attractors of multi-field inflation \cite{Bjorkmo:2019qno,Christodoulidis:2023eiw}. Here we use the same method for anisotropic inflation for $\mathcal{I}\sin{\theta}\gg 1$.

The conjugate momentum of the perturbations are given by
\begin{align}
  \pi_{\varphi}\equiv\frac{\partial\mathcal{L}}{\partial\dot{\varphi}^*}=\frac{a^3}{2}\dot{\varphi}, \ \ \ \ \ \pi_{\mathcal{A}}\equiv\frac{\partial\mathcal{L}}{\partial\dot{\mathcal{A}}^*}=\frac{a^3}{2}\left(\dot{\mathcal{A}}-4\mathcal{I}\sin{\theta}H\varphi\right).
\end{align}
After defining $p_{\varphi}\equiv2\pi_{\varphi}/a^3H$, $p_{\mathcal{A}}\equiv2\pi_{\mathcal{A}}/a^3H$, $x\equiv-k\eta$\, and using the e-folding number as the time variable, the coupled system can be written as $X_{,x}=J_{0}X+J_{1}X$, where
\begin{align}
  J_{0}=\left(\
  \begin{matrix}
  0 & 1 & 4\Omega & 0 &\\
  -x^2 & 0 & 0 & 0 &\\
  0 & 0 & 0 & 0 &\\
  0 & -4\Omega & -(x^2+8\mathcal{I}^2) & 0&
\end{matrix}
\right), \ \ \ \ \ J_{1}=\left(\
  \begin{matrix}
  0 & 0 & 0 & 0 &\\
  0 & -3 & -12\Omega & 0 &\\
  0 & 0 & 0 & 0 &\\
  -12\Omega & 0 & 0 & -3&
\end{matrix}
\right),
\end{align}
$X=(\mathcal{A},p_{\mathcal{A}},\varphi,p_{\varphi})^{T}$ and $\Omega=\mathcal{I}\sin{\theta}$. The matrix $J_{0}$ is $0$-$th$ order while $J_{1}$ is $1$-$st$ order. From the EFT of $\mathcal{A}$ we have known that the solutions are symmetric for $\Omega\to-\Omega$, so here we only assume that $\Omega>0$. We are also interested in the enhanced modes, hence we have $\sqrt{2}\Omega>\mathcal{I}$.  The eigenvalues of $J_{0}$ are $\pm\lambda_{\text{im}(\text{p})}$ with
\begin{align}
    \lambda_{\text{im}}&=i\sqrt{4\mathcal{I}^2+x^2+4\sqrt{\mathcal{I}^4+x^2\Omega^2}},\nonumber\\
    \lambda_{\text{p}}&=\sqrt{-4\mathcal{I}^2-x^2+4\sqrt{\mathcal{I}^4+x^2\Omega^2}}.
\end{align}
To ensure that $\lambda_{\text{p}}$ is positive, we need $x^2<-m_{\varphi}^2=8(2\Omega^2-\mathcal{I}^2)$.

To find the solution $X$ in the first order, we first diagonalize the matrix $J_{0}$ as $J_{\text{D}}=U^{-1}J_{0}U$. Then the equations of this system can be rewritten as the rotated vector $Y\equiv U^{-1}X$
\begin{align}
  Y_{,N}=J_{\text{D}}Y+\left(U^{-1}J_{1}U-U^{-1}U_{,N}\right)Y
\end{align}
The first term is $0$-$th$ order while the second one is $1$-$st$ order. In general, $U$ is not unique. To find the solutions up to corrections of the Hubble friction, $U$ can be chosen as the one that $U^{-1}J_{1}U-U^{-1}U_{,N}$ has only non-diagonal elements. To do this, one can first obtain $U$ from the eigenvectors $V^{(\alpha)}$ of $J_{0}$,: $U^{-1}_{ij}V^{(\alpha)}_{j}=\delta^{\alpha}_{\ i}$. We get
\begin{align}\label{U}
    U=\left(\
  \begin{matrix}
  \frac{\mathcal{I}^2-\sqrt{\mathcal{I}^4+x^2\Omega^2}}{x^2\Omega^2} & \frac{\mathcal{I}^2-\sqrt{\mathcal{I}^4+x^2\Omega^2}}{x^2\Omega^2} & \frac{\mathcal{I}^2+\sqrt{\mathcal{I}^4+x^2\Omega^2}}{x^2\Omega^2} & \frac{\mathcal{I}^2+\sqrt{\mathcal{I}^4+x^2\Omega^2}}{x^2\Omega^2} &\\
  \frac{\mathcal{I}^2-\sqrt{\mathcal{I}^4+x^2\Omega^2}}{\lambda_{\text{im}}\Omega} & -\frac{\mathcal{I}^2-\sqrt{\mathcal{I}^4+x^2\Omega^2}}{\lambda_{\text{im}}\Omega} & \frac{\mathcal{I}^2+\sqrt{\mathcal{I}^4+x^2\Omega^2}}{\lambda_{\text{p}}\Omega} & -\frac{\mathcal{I}^2+\sqrt{\mathcal{I}^4+x^2\Omega^2}}{\lambda_{\text{p}}\Omega} &\\
  -\frac{1}{\lambda_{\text{im}}} & \frac{1}{\lambda_{\text{im}}} & -\frac{1}{\lambda_{\text{p}}} & \frac{1}{\lambda_{\text{p}}} &\\
  1 & 1 & 1 & 1&
\end{matrix}
\right)
\end{align}
Then we divide $U^{-1}J_{1}U-U^{-1}U_{,N}$ into diagonal and off-diagonal parts $U^{-1}J_{1}U-U^{-1}U_{,x}=P_{\text{D}}+P_{\text{od}}$. The transformation matrix we want to find, which is denoted as $\tilde{U}$, can be obtained by
\begin{align}
    \tilde{U}=U\text{exp}\left(\int dN P_{\text{D}}\right).
\end{align}
It's not difficult to see that multiplying a diagonal matrix $\int dN P_{\text{D}}$ do not change the diagonal matrix $J_{\text{D}}$. So one can solve the leading-order solutions of $Y_{,N}=J_{\text{D}}Y$, which can be written as the fundamental matrix
\begin{align}
    F_{Y}=\text{diag}\left(e^{-\int dN\lambda_{\text{im}}},e^{\int dN\lambda_{\text{im}}},e^{-\int dN\lambda_{\text{p}}},e^{-\int dN\lambda_{\text{p}}}\right),
\end{align}
and subsequently obtain the fundamental matrix of $X$ by $F_{X}=\tilde{U}F_{Y}$.

From (\ref{U}) the diagonal matrix $ P_{\text{D}}$ is given by
\begin{align}
    P_{\text{D},11}=P_{\text{D},22}=-\frac{3}{2}+x^2\bigg[&4\mathcal{I}^6+2\mathcal{I}^4\big(x^2-4\Omega^2+2\sqrt{\mathcal{I}^2+x^2\Omega^2}\big)+x^2\Omega^2\big(x^2+8\Omega^2+6\sqrt{\mathcal{I}^4+x^2\Omega^2}\big)\nonumber\\
    &-8\mathcal{I}^2\Omega^2\big(\sqrt{\mathcal{I}^4+x^2\Omega^2}\big)-\mathcal{I}^2x^2\big(4\Omega^2+\sqrt{\mathcal{I}^4+x^2\Omega^2}\big)\bigg]\bigg{/}2\big(\mathcal{I}^4+x^2\Omega^2\big)\lambda_{\text{im}}^2,\nonumber\\
    P_{\text{D},33}=P_{\text{D},44}=-\frac{3}{2}+x^2\bigg[&2\mathcal{I}^4+\Omega^2\big(x^2-2\sqrt{\mathcal{I}^4+x^2\Omega^2}\big)+\mathcal{I}^2\big(-4\Omega^2+\sqrt{\mathcal{I}^4+x^2\Omega^2}\big)\bigg]\bigg{/}(\mathcal{I}^4+x^2\Omega^2)\lambda_{\text{p}}^2.
\end{align}
Then we get
\begin{align}
     \left(e^{\int dN P_{\text{D}}}\right)_{11}=\left(e^{\int dN P_{\text{D}}}\right)_{22}&=2^{1/4}\Omega^{-1/2}x\big(\sqrt{\mathcal{I}^4+x^2\Omega^2}+\mathcal{I}^2\big)f_{+}(x)\nonumber\\
     \left(e^{\int dN P_{\text{D}}}\right)_{33}=\left(e^{\int dN P_{\text{D}}}\right)_{44}&=2^{1/4}\Omega^{-1/2}x\big(\sqrt{\mathcal{I}^4+x^2\Omega^2}-\mathcal{I}^2\big)f_{-}(x).
\end{align}
where
\begin{align}
    f_{\pm}(x)&=\left[\frac{\big(\mathcal{I}^2\mp\sqrt{\mathcal{I}^4+x^2\Omega^2}\big)\big(\mathcal{I}^2-4\Omega^2\mp\sqrt{\mathcal{I}^4+x^2\Omega^2}\big)}{2\big(\mathcal{I}^4+x^2\Omega^2\big)}\right]^{1/4}\nonumber\\
    &=\left[1\pm\frac{2\Omega^2-\mathcal{I}^2}{\sqrt{\mathcal{I}^4+x^2\Omega^2}}-\frac{\Omega^2\big(x^2+4\mathcal{I}^2\big)}{2\big(\mathcal{I}^4+x^2\Omega^2\big)}\right]^{1/4}
\end{align}
Finally, we obtain the transformation matrix
\begin{align}
    \tilde{U}_{11}=\tilde{U}_{12}=-xf_{+}(x),\ \ \ \ \ \tilde{U}_{13}=\tilde{U}_{14}=xf_{-}(x)
\end{align}
and the solution of the perturbation $\mathcal{A}$
\begin{align}
    \mathcal{A}=xf_{+}(x)\left(c_1 e^{-\int dN\lambda_{\text{im}}}+c_2 e^{\int dN\lambda_{\text{im}}}\right)+xf_{-}(x)\left(c_3 e^{-\int dN\lambda_{\text{p}}}+c_4 e^{\int dN\lambda_{\text{p}}}\right)
\end{align}
where the coefficients are $c_i\sim\mathcal{O}(1)$. 
\begin{figure}[tbp]
\centering
\includegraphics[scale=0.61]{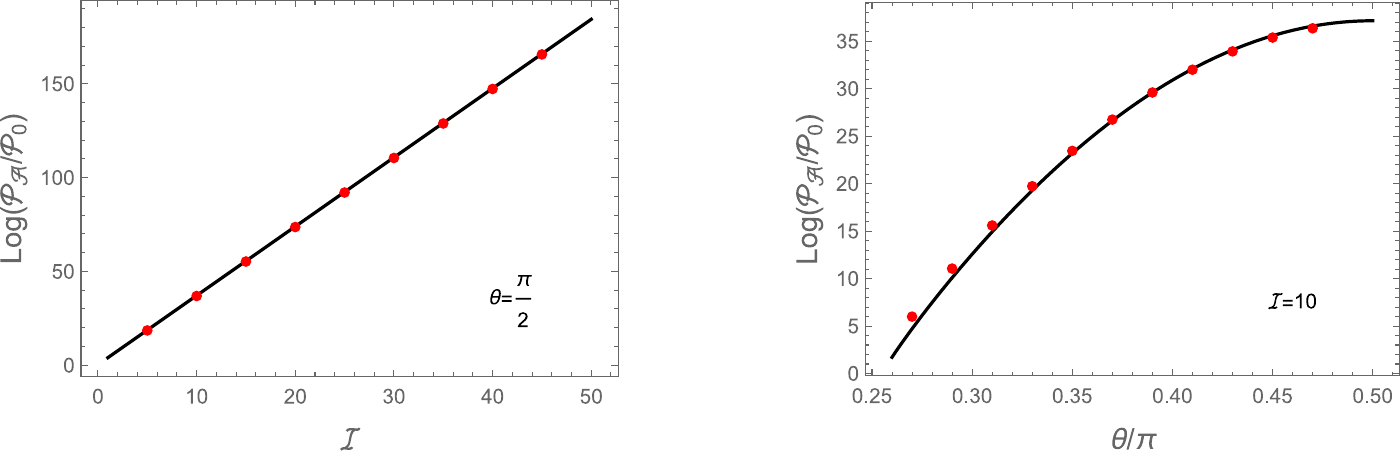}
\caption{\label{fig:EntropicF} Power spectrum $\mathcal{P}_{\mathcal{A}}/\mathcal{P}_0$. Solid curve is the analytical approximation (\ref{Pa}), while red dots are the numerical results by solving the equations.}
\end{figure}

The solution nere horizon-crossing is dominated by the exponential mode $e^{\int dN\lambda_{\text{p}}}$. After horizon-crossing $x\simeq 1$, this mode is frozen. Hence we can calculate the power spectrum as
\begin{align}
    \mathcal{P}_{\mathcal{A}}\simeq\mathcal{P}_0|f_{-}(x)|^2\text{exp}\left[2\int_{|m_{\varphi}|/H}^{x} dx'\left(-\frac{1}{x'}\right)\lambda_{\text{p}}\right]\Bigg{|}_{x=1},
\end{align}
where $\mathcal{P}_0\simeq x^2$ for $x\simeq 1$. The exponential factor can be calculated as
\begin{align}
    I(x)\equiv \int_{|m_{\varphi}|/H}^{x} dx'\left(-\frac{1}{x'}\right)\lambda_{\text{p}}=\frac{\mathcal{I}^2}{2\Omega}\Bigg[&-2\sqrt{1-b+by+y^2}+b\bigg(\frac{\pi}{2}+\arctan{\frac{b-2y}{2\sqrt{1-b+by-y^2}}}\bigg)\nonumber\\
    &-\sqrt{2b}\bigg(\frac{\pi}{2}-\arctan{\frac{2(1+y)+b(y-3)}{2\sqrt{2b}\sqrt{1-bc+by-y^2}}}\bigg)\Bigg],
\end{align}
where $b=4\Omega^2/\mathcal{I}^2$ and $\mathcal{I}^4y^2=\mathcal{I}^4+x^2\Omega^2$. It is convenient to take the limit $x\to0$ in the power spectrum. then we have 
\begin{align}
    I(x=0)=2\pi\left(\sin\theta-\frac{\sqrt{2}}{2}\right)\mathcal{I}
\end{align}
and the power spectrum of $\mathcal{A}$ can be approximately calculated by
\begin{align}\label{Pa}
    \mathcal{P}_{\mathcal{A}}\simeq\mathcal{P}_0\left|2-4\sin^2\theta\right|^{1/2}e^{2I(x=0)}.
\end{align}

\section{Power spectrum with Gaussian $\mathcal{I}$}
\begin{figure}[tbp]
\centering
\includegraphics[scale=0.55]{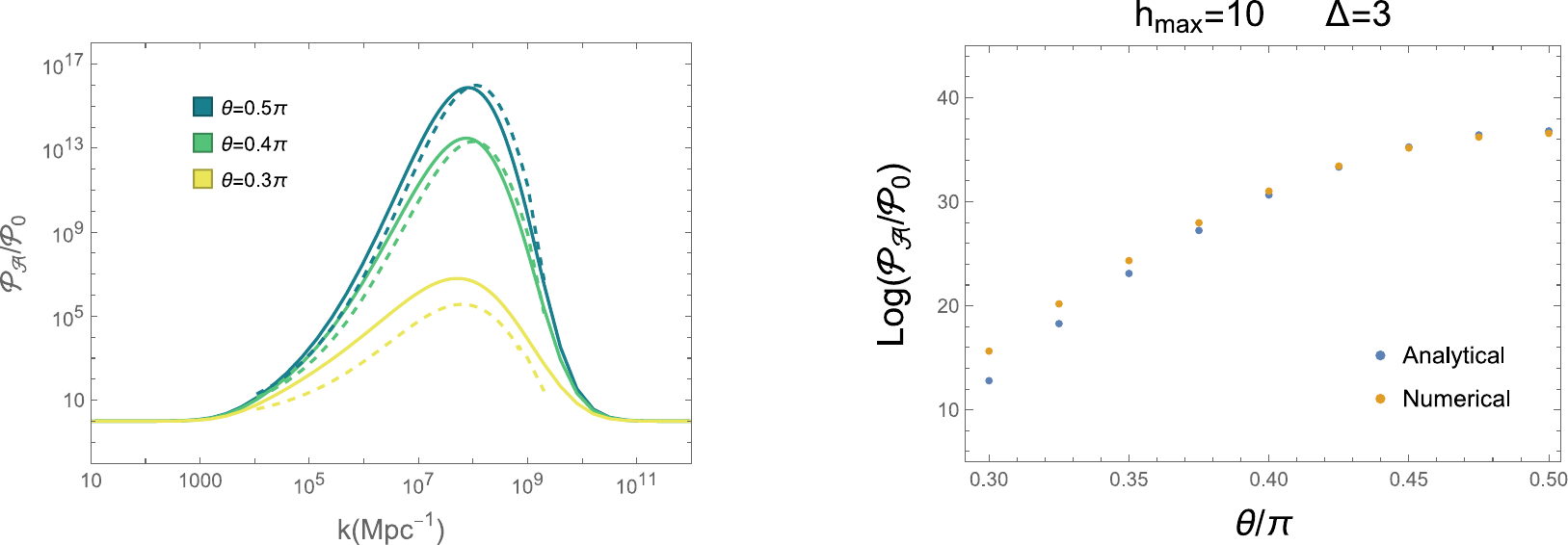}
\caption{\label{fig:PAAN} (Left): Power spectrum $\mathcal{P}_{\mathcal{A}}/\mathcal{P}_0$ for $\theta=0.3\pi,0.4\pi,0.5\pi$. Solid curves are the analytical results, while dashed curve are the numerical results by solving the equations. (Right): Comparison of Analytical and Numerical Results for the Peak of the Power Spectrum.}
\end{figure}

Here we use gaussian type of $\mathcal{I}(N)$ to calculate the power spectrum $\mathcal{P}_{\mathcal{A}}$. For numerical calculations, we replace $\mathcal{I}$ in the quadratic action by
\begin{align}
    \mathcal{I}(N)=\mathcal{I}_{\text{max}} e^{-(N-N_f)^2/(2\Delta N^2)},
\end{align}
and then derive the equations of motion of $\mathcal{A}$ and $\varphi$. The numerical results of the power spectrum of $\mathcal{A}$ are given by solving these coupled equations. We also calculated the power spectrum of $\mathcal{A}$ analytically by replacing $\mathcal{I}$ by gaussian $\mathcal{I}$ in (\ref{Pa}). Then $\mathcal{P}_{\mathcal{A}}$ is given by
\begin{align}
    \mathcal{P}_{\mathcal{A}}=\mathcal{P}_0\left|2-4\sin^2\theta\right|^{1/2}e^{4\pi\left|\sin{\theta}-\frac{\sqrt{2}}{2}\right|\mathcal{I}}\Big{|}_{N_v(k)}
\end{align}
where $N_v(k)$ is given by $p(N_v)=|m_{\varphi}(N_v)|$. We show the comparison between numerical and analytical results in Fig. \ref{fig:PAAN}. In the right panel of the figure, we compare the peak amplitudes of the power spectra for different values of $\theta$. It can be seen that for $\theta$ close to $\pi/2$, the numerical and analytical results agree well. However, as $\theta$ approaches $\pi/4$ – that is, when the peak has been sufficiently exponentially suppressed – the discrepancy between the two results becomes noticeable.

\section{Curvature perturbation of an ansiotropic universe}\label{app:curvature}

We use the ADM formalism for the metric of an anisotropic spacetime. The spacetime is split into the temporal parts and the spatial hypersurface. The metric is written as
\begin{align}
    &g_{00}=-N^2+N^aN_a,\ \ \ \ g_{0a}=N_a,\ \ \ \ g_{ab}=h_{ab},\nonumber\\
    &g^{00}=-N^2,\ \ \ \ g^{0a}=N^{-2}N^a,\ \ \ \ g^{ab}=h^{ab}-N^{-2}N^aN^b.
\end{align}
The normal vector of the hypersurface is 
\begin{align}
    n_0=-N,\ \ \ \ n_a=0,\ \ \ \ n^0=N^{-1},\ \ \ \ n^a=-N^{-1}N^a.
\end{align}
We use Latin script to indicate the spatial index $a,b,c,...=1,2,3$. Then the fluid quantities can be defined as
\begin{align}
    E\equiv n_{\alpha}n_{\beta}T^{\alpha\beta},\ \ \ \ J_{a}\equiv-n_{\beta}T^{\beta}_{\ a},\ \ \ \ S_{ab}\equiv T_{ab},\ \ \ \ S\equiv h^{ab}S_{ab}.
\end{align}
We use Greek script to indicate $\alpha,\beta,\gamma,...=0,1,2,3$. Here, $E$ is the energy density, $S/3$ is the pressure and $J_a$ is the energy flux of the fluids.

Now we consider a perturbed Bianchi type-I universe whose metric in the ADM formalism can be written as
\begin{align}
    N=a(1+A),\ \ \ \ N_{a}=a^2\tilde{B}_{a},\ \ \ \ h_{ab}=a^2(\gamma_{ab}+C_{ab}),
\end{align}
where $\gamma_{ab}$ is a $3\times3$ diagonal matrix. If $\gamma_{ab}=\delta_{ab}$,
the spacetime reduces to isotropy. Here we use the conformal time $x^0$ in the metric. One can also transform $x^0$ into the cosmological time $t$ by $dt=adx^0$. On the other hand, the perturbed quantities of fluids are written as
\begin{align}
    E=\rho+\delta\rho,\ \ \ \ S=3(p+\delta p),\ \ \ \ J_a=aQ_a.
\end{align}
The background and perturbed equations of these variables can be obtained directly. Here we define the Hubble variable as $H\equiv\dot{a}/a$ and only present some of these equations from \cite{Noh:1987vk}. 

The background equations are
\begin{align}
    \text{Energy constraint equation:}\ \ \ \ &3H^2-8\pi G\rho=-\frac{1}{8}\dot{\gamma}_{ab}\dot{\gamma}^{ab}.\label{bg1}\\
    \text{Raychaudhuri equation:}\ \ \ \ &3(\dot{H}+H^2)+4\pi G(\rho+3p)=\frac{1}{4}\dot{\gamma}_{ab}\dot{\gamma}^{ab}.\label{bg2}\\
    \text{Energy conservation equation:}\ \ \ \ &\dot{\rho}+3H(\rho+p)=-\frac{1}{2}\dot{\gamma}_{ab}\dot{\gamma}^{ab}\label{bg3}.
\end{align}
And the perturbation equations are
\begin{align}
    \text{Energy constraint equation:}\ \ \ \ &4H\delta K+16\pi G\delta\rho-\frac{1}{a^2}\left(C^{ab}_{\ \ ,ab}-\Delta C^a_{\ a}\right)\nonumber\\&=-\frac{1}{2}\dot{\gamma}_{ab}\dot{\gamma}^{ab}A+\frac{1}{a}\dot{\gamma}_{ab}\tilde{B}^{a|b}+\frac{1}{2}\dot{\gamma}^{ab}\gamma_{bc}\dot{C}^{c}_{\ a}.\label{ece}\\
    \text{Momentum constraint equation:}\ \ \ \ &\frac{2}{3}\delta K_{,a}+8\pi GaQ_a-\frac{1}{2a}\left(\Delta \tilde{B}_a+\frac{1}{3}\tilde{B}^{b}_{\ ,ab}\right)+\frac{1}{2}\left(\dot{C}^b_{\ a,b}-\frac{1}{3}\dot{C}^b_{\ b,a}\right)\nonumber\\
    &=\frac{1}{2}\dot{\gamma}_{ab}A^{|b}+\frac{1}{2}\dot{\gamma}^{bc}C_{ac,b}+\frac{1}{2}\dot{\gamma}_{ac}C^{bc}_{\ \ ,b}-\frac{1}{4}\dot{\gamma}_{ac}C^{b\ |c}_{\ b}+\frac{1}{4}\dot{\gamma}_{bc}C^{cb}_{\ \ ,a}.\label{mce}
\end{align}
Here $\delta K$ is the perturbation of the extrinsic curvature $\delta K=3aA+a^{-1}B^{a}_{\ ,a}-\dot{C}^{a}_{\ a}/2$, $\Delta=\gamma^{ab}\partial_a\partial_b$ and a vertical bar "$|$" indicates the covariant derivative based on $\gamma_{ab}$.

The form of perturbations of the metric in our configuration can be written as \cite{Emami:2013bk}

\begin{align}\label{pertg}
    \delta g_{\mu\nu}=\left(\
\begin{matrix}
-2a^2A & a^2\partial_1\beta & ab(\partial_iB+B_i) \\
  & -2a^2\psi_1 & ab\partial_1(\partial_i\gamma+\Gamma_i) \\
  &   & b^2(-2\psi_2\delta_{ij}+2E_{,ij}+E_{i,j}+E_{j,i})
\end{matrix}
\right),
\end{align}
where $\beta$, $B$ and $\gamma$ are scalar, while $B_i$ and $\Gamma_i$ are transverse vector  $\partial^iB_i=\partial^i\Gamma_i=0$. Here we denote $i,j=2,3$. Note that there are no tensor perturbations because they have been removed by the transverse and traceless conditions. In this ansatz of the Bianchi type I metric and the gauge-fixing, we have
\begin{align}
    \gamma_{11}=1,\ \ \ \ \gamma_{22}=\gamma_{33}=\frac{b^2}{a^2},\ \ \ \ \tilde{B}_a=\left(
    \begin{matrix}
    \partial_1\beta, \frac{b}{a}(\partial_iB+B_i)
    \end{matrix}
    \right),\ \ \ \  C_{ab}=\left(
\begin{matrix}
   0 & \frac{b}{a}\partial_1(\partial_i\gamma+\Gamma_i) \\
    & 0 \\
    & 0
\end{matrix}
\right),
\end{align}
All the terms in the r.h.s. of the equations above contain $\dot{\gamma}_{ab}$, which is computed as
\begin{align}
    \dot{\gamma}_{11}=0,\ \ \ \ \dot{\gamma}_{22}=\dot{\gamma}_{33}=\frac{2b^2}{a^3}\left(\frac{b'}{b}-\frac{a'}{a}\right)=\frac{2b^2}{a^2}H\epsilon_A.
\end{align}
Then, terms $\dot{\gamma}_{ab}\sim \epsilon$ and $\dot{\gamma}_{ab}\dot{\gamma}^{ab}\sim \epsilon^2$ in the r.h.s. of the equations.

Then we want to cancel the $\delta K$ in the equations to find a relation between $\delta\rho$ and $Q_a$. We first contract (\ref{mce}) by a opeator $\partial^a/\Delta$, and transform it into momentum space,
\begin{align}
    \frac{2}{3}\delta K-8\pi Ga\frac{ik^a}{k^2}Q_a+\mathcal{O}(\epsilon^2)-\frac{2ik^a}{3a}\tilde{B}_a+\frac{k^ak_b}{2k^2}\dot{C}^{b}_{\ a}+\mathcal{O}(\epsilon)=0.
\end{align}
We have collected all terms with $\dot{\gamma}_{ab}$ in $\mathcal{O}(\epsilon)$, and also used $\mathcal{O}(\epsilon^2)$ to indicate $\dot{\gamma}_{ab}\dot{\gamma}^{ab}A$ term. Combining with (\ref{ece}), we obtain
\begin{align}\label{rhoQ}
    4\pi G\delta\rho+12\pi GaH\frac{k^a}{k^2}Q_a+\mathcal{O}(\epsilon^2)=&-\frac{iHk^a}{a}\tilde{B}_a-\frac{k^2}{4a^2}\left(\frac{k_ak_b}{k^2}C^{ab}-C^{a}_{\ a}\right)\nonumber\\
    &+\frac{3H}{4}\left(\frac{k^ak_b}{k^2}\dot{C}^{b}_{\ a}-\frac{1}{3}\dot{C}^{a}_{\ a}\right)+\mathcal{O}(\epsilon).
\end{align}
In our ansatz of the metric, the r.h.s. can be computed as
\begin{align}
   \frac{k^2}{a^2}\tilde{\Psi}(\theta)\equiv\frac{k^2}{a^2}\left[H\left(a\beta\cos^2\theta+bB\sin^2\theta-\frac{3}{2}ab\dot{\gamma}\cos^2\theta\sin^2\theta\right)+\frac{k^2b}{2a}\gamma\cos^2\theta\sin^2\theta+\mathcal{O}(\epsilon)\right],
\end{align}
It is convenient to introduce $aQ_a\equiv(\rho+p)V_a$, where $V_a$ is the total velocity of the fluids. After using the background equations (\ref{bg1}), (\ref{bg2}) and (\ref{bg3}), equation (\ref{rhoQ}) can be written as
\begin{align}
    3\dot{H}\left[H\frac{\delta\rho}{\dot{\rho}}-H\frac{ik^a}{k^2}V_a+\mathcal{O}(\epsilon)\right]=\frac{k^2}{a^2}\tilde{\Psi}(\theta),
\end{align}
where we have swept all terms containing $\dot{\gamma}_{ab}$ in $\mathcal{O}(\epsilon)$, which is slow-roll suppressed compared to the other two terms. The r.h.s. tends to zero on superhorizon scales $k^2/a^2H^2\to0$.

To define a gauge-invariant curvature perturbation in a uniform-energy surface, we need to know the gauge transformation of $\delta \rho$ and the spatial perturbation. After decomposing $C_{ab}=C\gamma_{ab}+\bar{C}_{,ab}+2C^{(V)}_{(a,b)}+C^{(T)}_{ab}$, the gauge transformation of coordinates $x^{\alpha}\to x^{\alpha}+\xi^{\alpha}(x)$ results in transformations of $C$ and $\delta\rho$ as \cite{Noh:1987vk}
\begin{align}
    C\to C-\left(2H+\frac{\dot{\Delta}}{2\Delta}\right)a\xi^0,\ \ \ \ \delta\rho\to\delta\rho-\dot{\rho}a\xi^0,
\end{align}
where $\dot{\Delta}=\dot{\gamma}^{ab}\partial_a\partial_b$. Then the gauge-invariant curvature perturbation can be defined as
\begin{align}
    \zeta\equiv \frac{C}{2}-\left(H+\frac{\dot{\Delta}}{4\Delta}\right)\frac{\delta\rho}{\dot{\rho}}
\end{align}
In our ansatz of metric, $C=-2\psi_1$ and the second term in the brackets is sub-dominated due to $\dot{\gamma}_{ab}\sim\epsilon$. Then one can also define a variable as
\begin{align}
    \mathcal{R}\equiv -\frac{C}{2}+H\frac{ik^a}{k^2}V_a.
\end{align}
It reduces to the comoving curvature perturbation in the isotropic universe. On superhorizon scales, $\zeta$ and $\mathcal{R}$ differ only by a slow-roll correction $6\dot{H}\left[R+\zeta+\mathcal{O}(\epsilon)\right]\simeq0$. Hence, we can use $\mathcal{R}$ to calculate the leading-order contribution of the curvature perturbation $\zeta$. For gauge-fixing in our main text, $C=0$.

In the model we consider in the main text, the curvature perturbation can be calculated as
\begin{align}
    \mathcal{R}=H\frac{ik^a}{k^2}\frac{a}{\rho+P}\sum_{\gamma}Q_{a}^{(\gamma)},
\end{align}
where $\gamma=\phi,A$ represent two fields in the model and $\rho+P=\dot{\phi}^2+2f^2\dot{A}^2/3a^2$. $Q_a^{(\gamma)}$ can be computed directly from $Q_a=T^{0}_{\ a}/a$ as
\begin{align}
    a\frac{ik^a}{k^2}Q_{a}^{(\phi)}=\dot{\phi}\varphi,\ \ \ \ \ \ \ a\frac{ik^a}{k^2}Q_{a}^{(A)}=\frac{f\dot{A}\sin\theta}{a}\mathcal{A}.
\end{align}
Then we obtain
\begin{align}
    \mathcal{R}=H\frac{\dot{\phi}\varphi+\frac{f\dot{A}\sin\theta}{a}\mathcal{A}}{\dot{\phi}^2+\frac{2f^2\dot{A}^2}{3a^2}}=\frac{1}{\sqrt{2\epsilon_H}M_{\text{pl}}H}\frac{-\varphi+\mathcal{I}\sin\theta\mathcal{A}}{\sqrt{1+2\mathcal{I}^2/3}}.
\end{align}

\bibliographystyle{apsrev4-1}
\bibliography{Aniso_PBH_ref.bib}